\documentclass[aps,prb,groupedaddress,amsmath,amssymb,twocolumn]{revtex4}
    \usepackage{dcolumn}
    \usepackage{bm}
    \usepackage{graphicx}
    \usepackage{dcolumn}

\newcommand\beq{\begin{equation}}
\newcommand\eeq{\end{equation}}
\newcommand\beqa{\begin{eqnarray}}
\newcommand\eeqa{\end{eqnarray}}
\newcommand{\nn}{\nonumber\\}
\begin{document}
%\preprint{}

\title{Percus--Yevick theory for the structural properties of the seven-dimensional hard-sphere
fluid }


\author{Miguel Robles}
\email[]{mrp@cie.unam.mx}
\author{Mariano L\'opez de Haro}
\email[]{malopez@servidor.unam.mx} \affiliation{Centro de
Investigaci\'on en Energ\'{\i}a, UNAM, Temixco, Morelos 62580,
M{e}xico}
\author{Andr\'es Santos}
\email[]{andres@unex.es}
\homepage[]{http://www.unex.es/fisteor/andres/}
%\thanks{}
\affiliation{Departamento de F\'{\i}sica, Universidad de
Extremadura, E-06071 Badajoz, Spain}


\date{\today}

\begin{abstract}
The direct correlation function and the (static) structure factor
for a seven-dimensional hard-sphere fluid are considered. Analytical
results for these quantities are derived within the Percus--Yevick
theory
\end{abstract}

\maketitle

Interest in studying hard-core fluids in different dimensionalities
is tied to the notion that, since these systems share some
characteristic features such as the existence of a freezing
transition, by considering a higher dimension [in which the
mathematics may sometimes become simpler (see for instance Ref.\
\onlinecite{FP99})] one may gain insight into common phenomenology
which is either untractable or rather difficult in two or three
dimensions. Therefore it is not surprising that many studies have
been devoted to hard-core fluids in dimensions higher than three,
including fairly recent ones, mostly involving virial coefficients
and/or equations of state (for a non exhaustive but hopefully
representative list see Refs.\
\onlinecite{FI81,L84,LB82,RHS04,LB06}). On the other hand, the
importance of the Percus--Yevick (PY) theory\cite{PY} for the
structural and thermodynamic properties of liquids rests on the fact
that it is exactly solvable in the case of hard-core systems in odd
dimensions.\cite{FI81,L84} In fact, the direct correlation function
$c(r)$ and the static structure factor $S(q)$ have been derived
analytically in one, three, and five dimensions (see Ref.\
\onlinecite{L84} and references therein). The major aim of this Note
is to provide the explicit expressions for the functions $c(r)$ and
 $S(q)$ of the 7D hard-sphere fluid in the PY
theory. This work extends and complements our previous
paper\cite{RHS04} in which we concentrated on the thermodynamic
properties and the virial coefficients of the same system, also
within the PY approximation, and on the comparison of some of these
results with our own simulation data. It must be emphasized that the
highest dimensionality for which completely analytical results may
be derived from the PY theory is precisely $d=7$. Beyond that
dimensionality, numerical work is required at one stage or another.

\begin{figure}[h]
\includegraphics[width=0.8\columnwidth]{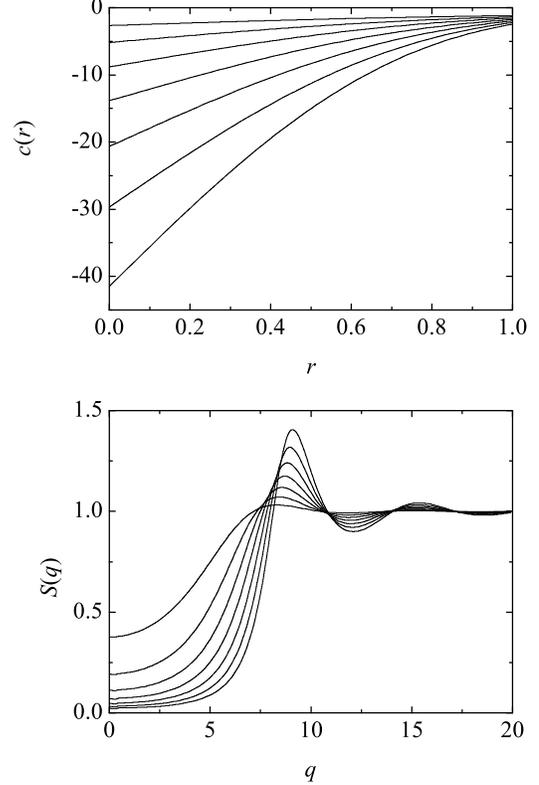}
\caption{Plot of the direct correlation function $c(r)$ (top panel)
and of the structure factor $S(q)$ (bottom panel) of the 7D fluid of
hard hyperspheres in the PY theory. The packing fractions
corresponding to the curves are, from top to bottom in the left end,
$\eta=0.01$, $0.02$, $0.03$, $0.04$, $0.05$, $0.06$, and
$0.07$.\label{fig1}}
\end{figure}

Following Ref.\ \onlinecite{L84}, the direct correlation function of
the 7D hard-sphere fluid  has, in the PY theory, the polynomial form
\begin{equation}
c(r)=\left(c_0+c_1 r+c_3 r^3+c_5 r^5+c_7 r^7\right)\Theta(1-r),
\label{1}
\end{equation}
where the diameter of a sphere is $\sigma=1$ and $\Theta(x)$ is
Heaviside's step function. The coefficients $c_i$ ($i=0,1,3,5,7$)
depend on the packing fraction $\eta=(\pi^3/840)\rho$ (where $\rho$
is the number density). After some algebra, one finds  the following
structure:
\begin{equation}
 c_i=\frac{1}{(1-\eta)^4}\sum_{j=0}^{4}t_{ij}\left[Q^{(0)}\right]^j.
 \label{2}
\end{equation}
Here, $Q^{(0)}$ is the physical solution of a quartic
equation,\cite{RHS04} which reads
\begin{equation}
Q^{(0)}=\frac{1}{3360\eta(1-\eta)}\left({3-10\eta}-
{{T_1^{1/2}}-{T_2^{1/2}}}\right),
\end{equation}
where
\begin{equation}
 T_1=\frac{2}{3}(3+36\eta+10 \eta^2 )+\frac{1}{6}B,
\end{equation}
\begin{equation}
 T_2=2 {T_1}-\frac{1}{2}B-\left(3-114\eta-222
   \eta ^2-10 \eta ^3\right)T_1^{-1/2},
\end{equation}
\begin{equation}
 B=A^{1/3}+{ \left(3+36\eta+10 \eta^2\right)^2}{A^{-1/3}},
\end{equation}
\beq
A=-A_1+\sqrt{A_1^2-\left(3+36\eta+10
    \eta^2\right)^6},
\eeq
\beqa
A_1&=&459+43740 \eta-4644 \eta ^2+16524 \eta ^3-357948 \eta^4\nn
&&+160920 \eta ^5+23300 \eta ^6.
\eeqa
The coefficients $t_{ij}$ appearing in Eq.\ \eqref{2} are listed in
Table \ref{tij}. It can be checked that $c(1^-)=-g(1^+)$, as
expected from the continuity of the indirect correlation function
$\gamma(r)\equiv g(r)-c(r)-1$. As an illustration, the top panel of
Fig.~\ref{fig1} shows $c(r)$ for $\eta=0.01, 0.02,\ldots,0.07$.
 \squeezetable
\begin{table}
\caption{\label{tij}Expressions for the coefficients $t_{ij}$.}
\begin{ruledtabular}
\begin{tabular}{cc}
$j$&$t_{0j}$\\
\hline
    $0$&$-(1 + 6\eta)^2$\\
    $1$&$2688\eta(2+5\eta)(1+6\eta)$\\
   $2$&$-1053696\eta^2(9 + 50\eta + 60\eta^2)$\\
    $3$&$1011548160 \eta^3(3+4\eta)(2+5\eta)$\\
 $4$&$-141616742400\eta^4(3 + 4\eta)^2$\\
 \hline
 \hline
$j$&$t_{1j}$\\
\hline
    $0$&    $\frac{35}{16} \eta  (4+3 \eta )^2$\\
    $1$&    $-840 \eta  ( 4+3 \eta) \left(1+33 \eta +15 \eta^2\right)$\\
   $2$&    $6720 \eta  \left(12+1212 \eta+18783 \eta^2+17060 \eta ^3+
3750 \eta ^4 \right)$\\
    $3$&    $-45158400 \eta ^2 \left(3+36 \eta + 10 \eta ^2\right)
   \left(1+33 \eta +15 \eta ^2\right)$\\
 $4$&$6322176000 \eta ^3 \left(3+36\eta + 10 \eta ^2\right)^2$\\
 \hline
 \hline
$j$&$t_{3j}$\\
\hline
$0$&    $-\frac{35}{16} (20-13 \eta ) \left(12-38 \eta+ 33 \eta
^2\right)$\\
 $1$& $-420 \left(60-2759\eta+9163 \eta ^2-10975 \eta ^3+   3825
\eta ^4 \right)$\\
 $2$  &  $3360 \eta  \left(22041 -308370 \eta+812905 \eta
^2-882160\eta^3+ 273950 \eta ^4\right)$\\
 $3$&   $-22579200 \eta ^2 \left(3147 -19924 \eta +38685 \eta
^2-36360 \eta ^3+ 9650 \eta ^4 \right)$\\
$4$&
 $6322176000 \eta ^3 \left(3144-12576\eta +17541 \eta ^2-13360 \eta^3+ 2850 \eta ^4
 \right)$\\
 \hline
 \hline
$j$&$t_{5j}$\\
\hline
$0$&
 $\frac{21}{16} \left(1320-5107 \eta +6282 \eta ^2- 2446 \eta
^3 \right)$\\
 $1$&
    $ 252 \left(330 -15179 \eta+51509 \eta ^2 -59291 \eta^3+ 21945
\eta^4 \right)$\\
 $2$&
    $-14112 \eta  \left(17325-242482 \eta+661205\eta ^2-674260
\eta ^3+ 226550 \eta^4\right)$\\
 $3$&
    $94832640 \eta ^2 \left(2475-15664 \eta+31953 \eta ^2-27500
\eta ^3  +  8050\eta ^4 \right)$\\
$4$&$-26553139200\eta ^3 \left(2475-9888 \eta+14702 \eta
^2-10032 \eta ^3+ 2400 \eta ^4 \right)$\\
 \hline
 \hline
$j$&$t_{7j}$\\
\hline
    $0$&$-\frac{1}{16}\left(20592-79789 \eta+97872 \eta ^2-    38430
\eta ^3\right)$\\
    $1$&$-24 \left(2574-118404 \eta+402761 \eta ^2-461341 \eta^3+
172695 \eta^4 \right)$\\
    $2$&$47040 \eta  \left(3861 -54054 \eta +147942 \eta ^2-149642\eta
^3+  51060\eta ^4\right)$\\
    $3$&$-45158400 \eta ^2 \left(3861-24453 \eta+50151 \eta ^2-42632
\eta ^3+12730 \eta ^4 \right)$\\
    $4$&$6322176000 \eta ^3 \left(7722-30888 \eta
+46269 \eta ^2-31056 \eta ^3+ 7610 \eta ^4\right)$
\end{tabular}
\end{ruledtabular}
\end{table}

Now we turn to the structure factor $S(q)$. According to the
Ornstein--Zernike equation, $S(q)$ is related to the Fourier
transform $\widetilde{c}(q)$ of  $c(r)$ by
\begin{equation}
 S(q)=\frac{1}{1-\rho \widetilde c(q)}.
\label{S(q)}
\end{equation}
In 7D, the Fourier transform is\cite{L84}
\begin{equation}
 \widetilde{c}(q)=(2\pi)^{7/2}q^{-5/2}\int_0 ^\infty dr\, r^{7/2} J_{5/2}(qr)
 c(r),
 \label{3}
\end{equation}
where $J_{5/2}(z)=\sqrt{2/\pi}z^{-5/2}\left[(z^2-3)\sin z-3z\cos
z\right]$ is a Bessel function of the first kind. Inserting Eq.\
\eqref{1} into Eq.\ \eqref{3} and performing the integrals one gets
\beq
 \widetilde{c}(q)=c_0 \psi_0(q)+c_1 \psi_1(q)+c_3 \psi_3(q)+c_5\psi_5(q)+c_7
 \psi_7(q),
\label{4}
\eeq
where
\begin{equation}
\psi_i(q)=16\pi^3 q^{-(7+i)}\left[\mathcal{C}_i(q)
   \cos q   -\mathcal{S}_i(q) \sin
   q -\mathcal{C}_i(0)
\right].
\end{equation}
The functions $\mathcal{C}_i(q)$ are listed in Table \ref{psi_i},
while the functions  $\mathcal{S}_i(q)$ are
\beq
\mathcal{S}_i(q)=\frac{d}{dq} \mathcal{C}_i(q)+3q^{2+i}.
\eeq
Insertion of Eq.\ \eqref{4} into Eq.\ \eqref{S(q)} gives the
structure factor. This function is plotted in the bottom panel of
Fig.~\ref{fig1}.
\begin{table}
\caption{\label{psi_i}Expressions for the functions
$\mathcal{C}_i(q)$.}
\begin{ruledtabular}
\begin{tabular}{cc}
$i$&$\mathcal{C}_i(q)$\\
\hline
    $0$&$q \left(q^2-15\right) $\\
    $1$&$q^4-24
   q^2+48$\\
   $3$&$q^6-48q^4+576q^2-1152$\\
    $5$&$q^8-80 q^6+2400q^4-28800
q^2+57600$\\
    $7$&$ q^{10}-120q^8+6720 q^6-201600 q^4
+2419200 q^2 -4838400$
\end{tabular}
\end{ruledtabular}
\end{table}

In view of the rather satisfactory agreement that has been observed
between the analytical PY results for the compressibility factor of
the 7D hard-sphere fluid and the simulation data, \cite{RHS04,LB06}
one wonders whether a similar agreement will hold for the structural
properties. That this is indeed the case has been recently confirmed
by Bishop and his coworkers,\cite{Bishop} who have found a very good
agreement between their simulation results and the ones computed
using our analytical formulae.

\acknowledgments

This work has been partially supported by DGAPA-UNAM under project
IN-110406. One of the authors (A.S.) acknowledges the financial
support of Ministerio de Educaci\'on y Ciencia (Spain) through
Grant No. FIS2004-01399 (partially financed by FEDER funds).

\end{document}